\documentstyle[twocolumn,prl,psfig,aps]{revtex}
\def\x{\mbox{\bf x}}
\begin{document}
\title{Statistical Test for Dynamical Nonstationarity in Observed
Time-Series Data}
\author{Matthew B. Kennel}
\address{Engineering Technology Division, MS 8088,\\
Oak Ridge National Laboratory \\
Oak Ridge, TN 37831-8088\\
{\tt kennel@msr.epm.ornl.gov}}
\maketitle
\begin{abstract} Information in the time distribution of points in a
state space reconstructed from observed data yields a test for
``nonstationarity''.  Framed in terms of a statistical hypothesis
test, this numerical algorithm can discern whether some underlying
slow changes in parameters have taken place.  The method examines a
fundamental object in nonlinear dynamics, the geometry of orbits in
state space, with corrections to
overcome difficulties in real dynamical data which cause naive
statistics to fail.
\end{abstract}
Since the discovery of the time-delay embedding for state-space
reconstruction\cite{packard80,takens,embedology} a significant effort
has been devoted to the development of techniques to extract
information in observed time-series data from a geometrical, dynamical
viewpoint.  Underlying nearly all of these techniques (\cite{review}
is a review) is an assumption of {\em stationarity}: the dynamical
process, and hence the geometrical attractor containing the orbits,
has not changed on long time scales the order of the length of the
dataset.  If not true, there may be significant behavior on timescales
longer than may be reliably resolved with the given data, or perhaps,
experimental parameters, presumed fixed, have actually changed during
the run.

Despite its nearly universal assumption, there is little previous
literature on reliably testing for stationarity in physical
situations. This work demonstrates a statistic and associated
hypothesis test which sensitively detects nonstationary behavior given
broadband and potentially chaotic data.  A stationary dataset is
presumed to to be sufficiently long to trace out a good approximation
to the invariant measure. The algorithm described in this work
quantifies ``how much has the invariant measure, as inferred from the
observations, changed over long timescales'' and whether ``this change
is statistically significant.''

One could imagine measuring any number of simple statistics, such as
the mean or standard deviation, from the two halves of the time
series, and constructing a hypothesis test based on their presumed
equality, but such techniques are not particularly good.  First, the
statistic is arbitrary and not related to the natural geometrical
properties of the attractor, which we presume is the interesting
object when analyzing chaos and other dynamical data.  If not the mean
or first moment, one could have chosen the average of, say, the
third Legendre polynomial of the orbit point dotted into some
arbitrary vector, {\em et cetera}, until one found the answer one
wanted.  Unless the particular statistic estimates a parameter deemed
physically or dynamically important, such arbitrary choices are not
particularly enlightening, and their power against various sorts of
nonstationarity vary greatly.  Second, naively applying procedures
greatly overestimates the significance of differences: observed
dynamical data are far from uncorrelated yet the simple, classical
statistical estimations of confidence rely on the notion of
independent observations.  For example, measuring empirical means of
first and second halves of a chaotic dataset and performing the
classical $t$-test for their equality will quite often spuriously (and
vehemently) reject the null hypothesis of stationarity, even when the
data come from clean stationary experiments or well-known simple
models such as the Lorenz attractor.  Such methods do not not reliably
diagnose the intuitive concept of dynamical stationarity that a typical
physicist would imagine.

The present work attempts to rectify these two issues. One, by
measuring a quantity related to geometrical properties in the full
state space, and two, by accounting for the temporal
dependence intrinsic in orbits from a continuous-time dynamical system.
Furthermore, the method does not require artificially partitioning the time
series into halves or other smaller time segments.

I sidestep {\em direct} estimations of the invariant measure from observed
data.  Counting points in boxes of state space, as used for
computing mutual information~\cite{fraser}, for instance, introduces
potentially problematic issues such as the arbitrary choice of box
size, quantization artifacts, and poor scaling with the
embedding dimension.  Kernel density estimators are computationally
intensive in higher dimensions and functionals or statistics on
such estimates may require difficult multi-dimensional integrals.
The formalism does not naturally offer clear tests for significance.

Instead, the solution adopted quantifies nonstationarity
using properties of {\em nearest neighbors} in state space.  Neighbor searching
is efficient and the estimates of consequent properties do not have a
{\em prima facie} exponential ``curse of dimensionality''.  Neighbor
statistics were used in \cite{fnn,fns} to determine minimum embedding
dimension for reconstruction, and to quantify predictability
of observed chaotic data \cite{casdagli91}.

%

As background motivation, suppose we have two empirical
probability distributions $\rho_1$ and $\rho_2$, the
measures in the first and second halves of the dataset.  One wonders
whether $\rho_1 = \rho_2$.  Rewrite as
\begin{eqnarray}
\rho_1 &=& \frac{\rho_1 + \rho_2}{2} + \frac{\rho_1 - \rho_2}{2} = \rho_0 +
\delta \rho \\
\rho_2 &=& \frac{\rho_1 + \rho_2}{2} - \frac{\rho_1 - \rho_2}{2} = \rho_0 -
\delta \rho
\end{eqnarray}

Given some $\x$ from the first half consider the
probability that its nearest neighbor $\x^{nn}$ is also in the same
half. Assuming $\rho(\x^{nn})\approx \rho(\x)$,
$p_{\mbox{same}} = \frac{\rho_1}{\rho_1 + \rho_2} = \frac{\rho_0
+\delta \rho}{2 \rho_0}$.  The expected proportion of matches is thus
$$
\int d\x \rho_1(\x) p_{\mbox{same}} = \langle \frac{(\rho_0 +
\delta\rho)^2}{2\rho_0}
\rangle_{\mbox{first half}}.$$

With the same argument for the second half we find the overall expectation of
seeing same-half matches is
\begin{eqnarray}
E(\mbox{same}) &=& \langle \frac {(\rho_0 - \delta\rho)^2 +
 (\rho_0 - \delta\rho)^2}{2 \rho_0} \rangle \nonumber \\
 &=& \langle \frac{\rho_0 ^2 + \delta \rho^2}{2 \rho_0}\rangle =
 \frac{1}{2} + \langle \frac{\delta\rho^2}{2 \rho_0} \rangle.
\end{eqnarray}
Nonstationarity, i.e. $\delta \rho \ne 0$, always increases
this quantity, meaning that neighbors in state space are especially
close in {\em time} when the distribution drifts over time.

The actual statistic feels the same effect but is more subtle:
one collects the distribution of $D \equiv |\Delta(\x)| \equiv
|(T(\x^{nn})-T(\x)|$ for all observed $\x$, where $T(\cdot)$ denotes
the time index of the point.  Nonstationarity induces an excess number
of small values of $D$ than otherwise expected.

Naively counting up the $D$ from all points and their nearest neighbors
does not render a successful algorithm.  As with computing
the correlation dimension \cite{jt91}, one must
exclude neighbors close in time because they are not independent of
the reference point.  If a prospective neighbor would result in
$D =  |T(\x^{nn}) - T(\x)| < W$, ignore it and continue
searching instead.  The interval $W$ is set to a characteristic
autocorrelation time, perhaps 3 to 5 times the first minimum of mutual
information.

Equally important, but less obvious, is accounting for serial
correlation of neighboring trajectories: iterates of nearest neighbors
often remain nearest neighbors, but this does not give {\em new}
information.  The present algorithm gathers multiple pairs of points and
their neighbors which share the same $\Delta$ into the same {\em strand}.  If
the $\Delta$ associated with $\x(i)$ is the same as that for $\x(i-k)$
for any $k\in [1,W]$ append $\x(i)$ and its nearest neighbor to the
strand associated with $\x(i-k)$.  Otherwise, start a new strand with
$\x(i)$ and its nearest neighbor with the as yet sole element.  Note
that elements of a single strand are not necessarily consecutive;
there can be a gap up to $W$ timesteps long, though this is rarely
realized in practice.  Allowing such gaps prevents noise from damaging
the proper accounting of neighbor correlations.  Strands have two
pieces, the ``reference'' section and the ``neighbor'' section, whose
underlying points are nearest neighbors to the points in the reference
section.

The final correction culls strands which share underlying points,
whether in the reference or neighbor part, because their information
is not completely independent.  If any pair of strands share any points,
we randomly delete one of the strands until no remaining strands
share any points. Without the corrections, the ``N'' used in
statistics is larger than it should be and would cause spurious null
violations.

\begin{figure}
\centerline{\psfig{file=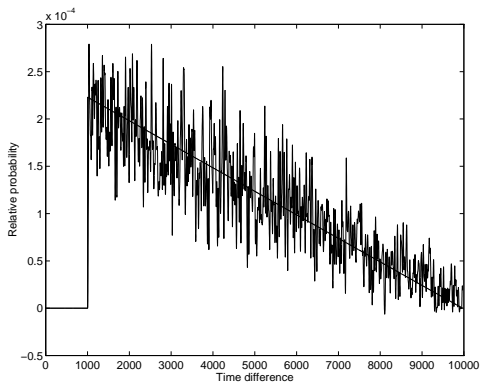,height=2in}}
\caption{An example of expected and observed probability density functions for
$D$ under the null hypothesis.}
\label{pdfs}
\end{figure}

We test the observed distribution of $D$ for the final set of strands
against the distribution expected under the null hypothesis of
stationarity.  The null assumes the time index of a neighbor is
independent from that of the reference.  For each observed strand,
we pretend that the neighboring portion could have started at any time
index in $[1,N]$ with equal probability, excluding the interval $W$
steps before the start and $W$ steps after the end of the reference
portion.  This generates an expected distribution of $D$ focused
around that one
strand; we repeat for all strands, generating the overall
expected distribution of $D$, and normalize when done.
This procedure takes computation time
$O(N_{\mbox{points}} \cdot N_{\mbox{strands}} )$ and so may be slow.
An approximation good for reasonably large $N$ is the
triangular shaped function derived by considering strands as points:
$$\begin{tabular}{cllr}
$p(D)$&$=$& $0,$   & $D \in [0,W]$ \\
$p(D)$&$=$& $M^{-1}\frac{N-D}{N-(W+1)},$ & $D \in [W+1,N-1] $
\end{tabular} $$
with $M$ chosen to normalize $p(D)$.  Figure~\ref{pdfs}
shows an example observed and expected distribution
as an illustration of the typical shape.

This picture suggests using the Kolmogorov-Smirnov test
on observed and expected $p(D)$.
In practice, that statistic turned out overly sensitive to medium time-scale
dynamical fluctuations observed even in stationary attractors.
Instead, a ``sign test'' provides an even simpler and effective test
which is most sensitive to the long timescale changes characteristic of
nonstationarity.  Denote the location at the median
of the expected distribution as $D^*$.  Then one counts the proportion of
actually observed strands with $D<D^*$,
$p = N_{\mbox{observed}} / N_{\mbox{strands}}$.  One expects $p_0 = 0.5$.
Under the null hypothesis,
\begin{equation}
z = \left( p-p_0 \right) \left( {N_{\mbox{strands}}}^{-1} p_0(1-p_0)
\right)^{-1/2} \label{zeq}
\end{equation}
is $N(0,1)$.  Thus if one observes $z > 2.36$ one rejects
the null at the 99\% confidence level.  Unlike a K-S test, this
sign test ensures that the violation be in the proper direction to
be caused by nonstationarity, which causes {\em large}
values of $p$ and hence $z$.  Significant, but negative values of $z$ suggest
important non-uniform neighbor time differences distinct from nonstationarity.
Strong low-frequency periodic behavior seems able to produce such results.

Any statistical inference is only as good as its assumptions, in this case,
that all strands are completely independent, and that in stationary systems
nearest neighbor time-delays are distributed completely uniformly.  This
is indeed true for unbiased stochastic draws from probability densities, but
is only an approximation for real dynamical systems.  In constrast to the
simple assumptions of classical statistics the diversity
of possible behaviors under nonlinearity makes it very difficult to construct
any interesting test where {\em chaos} is the null, something surrogate
data methods do not attempt. The present test does so by testing for one
specific aspect of dynamical systems and making an approximation
that empirically appears to reasonably good.  The main problem is that the
``level'' of the test, the frequency of finding $z>2.36$ under stationary
conditions, is not exactly calibrated to the supposed 1\%.  This does not
seem to be resolvable in general unless one had large amounts of the specific
attractor observed in stationary conditions which would
generate the actual distribution of $z$ in (\ref{zeq}) instead assuming
the normality. If the data were truly drawn
independently and randomly from probability functions $\rho_1$ and
$\rho_2$, the approximation is exact.  The value of this method is an
approach and approximation that works for many sorts of realistic
datasets without requiring a large database of previously observed
stationary orbits.

A computationally expensive but valuable confirmation protocol is to
examine the proportion of $z$ values which reject the null as computed
using randomly selected contiguous subsegments of lengths $N^* < N$ of
the original data: a ``poor-man's bootstrapping''.  With authentically
nonstationary behavior, this proportion rises steadily with $N^*$.
One may also examine the behavior with $N^*$ of $p$ averaged over
subsamples as well as its effective significance via (\ref{zeq}) to
check whether $p$ grows consistently large with $N^*$ and not just
wider than $N(0,1)$.

\begin{figure}
\centerline{\psfig{file=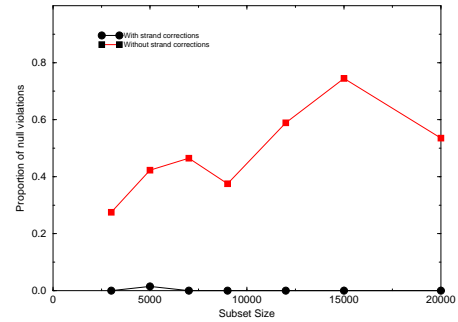,angle=270,height=2in}}
\caption{Proportion of rejections as a function of subsample size when
strand corrections are turned off and on.  Data set is from a stationary
low-dimensional chaotic circuit, but without strand corrections there is
frequent suprious rejection of the null.} \label{circuit}
\end{figure}

Figure~\ref{circuit} demonstrates the importance of the strand corrections.
The data come from an experimental nonlinear circuit used to investigate
synchronization and chaotic communications.  The dynamics are known to be
low-dimensional and the data clean.  In any useful sense the data are
quite stationary, yet the uncorrected statistic shows large violations,
as would naive tests found in statistical textbooks such as equality of means
or variances tried on first and second halves.  By constrast, the present
method shows no spurious null violations above the expected proportion.
Consistent with the bootstrapping analysis, the dataset {\em in toto}
violates the null without strand corrections but is consistent when those
corrections are reinstated.

\begin{figure}
\centerline{\psfig{file=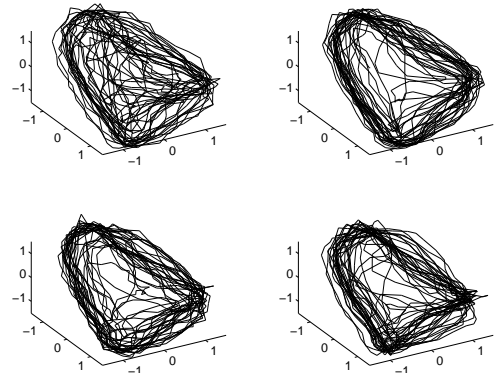,height=2in}}
\caption{Phase space plots of the differential pressure signal from a fluidized
bed reactor.  Three are from the same parameters, one is different.}
\label{fourbeds}
\end{figure}

\begin{figure}
\centerline{\psfig{file=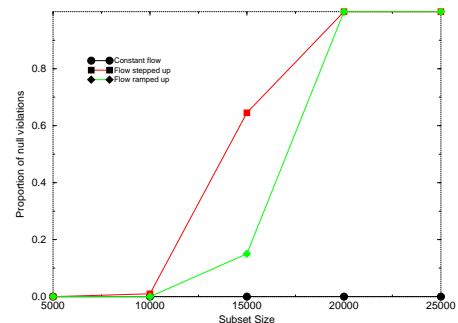,angle=270,height=2in}}
\caption{Proportion of rejections for stationary, step change and ramped
air flow.} \label{bedresult}
\end{figure}

The next example demonstrates a more concrete engineering application.
The data set was the pressure drop across a 15cm gap in a ``fluidized
bed reactor'', consisting of glass particles 2.7mm mean diameter in a
10cm diameter vertical cyclinder with air blown at constant flow from
the bottom, a small scale model of industrial chemical reactors.
For some external parameter regimes, the mass of particles
undegoes complex motion which appears to be a combination of
low-dimensional bulk dynamics and small-scale high-dimensional
turbulence of the individual particles.\cite{bedprl} The observed
variable was a pressure difference between two vertically separated
taps.  Figure~\ref{fourbeds} shows portions of time-delay embedding of
orbits sections of the dataset taken at the same experimental
parameters, and one when the air flow was boosted by 5\%.  The change
in the attractor is rather subtle and difficult to reliably diagnose
by eye.  The statistic distinguishes them easily:
Figure~\ref{bedresult} shows the bootstrapping result on three
datasets, one under stationary conditions, and one with a step change
to the higher flow, and one with a slow ramp to that same flow.  The
lower right figure (not the upper left) is from the data taken at a
different flow rate than the others.

The author has applied the method to quite a variety of data sets,
simulated and experimental, and it yields correct and appropriate
results in all cases found so far.  It is not sensitive to
reconstruction parameters and does not require the data to be known
{\em a priori} to be clean low-dimensionality: it is not clear whether
the fluidized bed datasets analyzed herein are better described as
``chaos'' or ``very noisy periodicity''.

There is a whole class of related statistics that use the same
neighbor principle.  Instead of $|(T(\x^{nn})-T(\x)|$ one may use the
distribution of any general function $f(\x^{nn},\x)$.  For instance,
$f(\cdot,\cdot)$ may be the ``indicator function'', yielding 1 if both
its arguments come from the same dataset and 0 otherwise.  This
provides a test for equivalence of the two data sets and can also
yield a distance measure.  The author has already done so to implement
a ``change-point-detector'' which accurately finds the particular
moment in time when some underlying parameter had changed, and the statistical
confidence of its authenticity.  Choosing
$f = |\sin(\Omega T(\x^{nn}) + \Phi)-\sin(\Omega T(\x) + \Phi)|$
yields a test for the presence, and statistical significance, of a slow
periodic modulation of the underlying attractor.  If one has measured some
other slowly varying signal $y(t)$, then the choice of $f =
|y(T(\x^{nn})) - y(T(\x))|$ provides a test whether there is any
statistically significant dynamical correlation betwen $y$ and the
pattern of orbits traced out by $\x$.  For instance, one might wish to
test the hypothesis that somehigh-frequency weather patterns in $\x$ is
significantly correlated with historical $CO_2$ levels.  These
variations, alternative stationarity algorithms based on the
correlation integral, as well as more extensive experimental results
will be investigated in the author's forthcoming research.

Isliker and Kurths~\cite{iskurths} propose testing the one dimensional
marginal distribution of the data for stationarity, but this ignores
dynamical time domain information, and their method does not appear to
account for serial correlation.  Brown et al\cite{brown} synchronize
empirical ODE models to time series, and propose using a long term
increase in deviation as a measure of nonstationarity.  This method
appears powerful and relies on non-trivial dynamical information but
requires clean low-dimensional data and does not provide an obvious
statistical test.

The author is indebted to many discussions with C. Stuart Daw, Charles
Finney, Ke Nguyen, and Martin Casdagli.  This research was supported
by the Department of Energy Distinguished Postdoctoral Fellow program.

\end{document}